# Statistical Comparison among Brain Networks with Popular Network Measurement Algorithms

Rakib Hassan Pran
M.Sc. in Applied Statistics with Network Analysis
International Laboratory for Applied Network Research
Research Departments of NRU HSE
National Research University Higher School of Economics
Moscow, Russia
Email: rakibhassanpra@gmail.com

*Abstract*—In this research, a number of popular network measurement algorithms have been applied to several brain networks (based on applicability of algorithms) for finding out statistical correlation among these popular network measurements which will help scientists to understand these popular network measurement algorithms and their applicability to brain networks. By analyzing the results of correlations among these network measurement algorithms, statistical comparison among selected brain networks has also been summarized. Besides that, to understand each brain network, the visualization of each brain network and each brain network's degree distribution histogram have been extrapolated. Six network measurement algorithms have been chosen to apply time to time on sixteen brain networks based on applicability of these network measurement algorithms and the results of these network measurements are put into a correlation method to show the relationship among these six network measurement algorithms for each brain network. At the end, the results of the correlations have been summarized to show the statistical comparison among these sixteen brain networks.

## I. Introduction

Networks are ubiquitous. From biology to technology, networks exist in every discipline of study [1]. To understand networks in any discipline, network analysis is fundamental while network analysis techniques are being developed rapidly day by day [2][3].

Besides applying network analysis techniques separately to understand a network for each technique, it is also necessary to understand the relationships among these techniques. Network measurement algorithms are pivotal parts of network analysis techniques [2]. Previously, few researches [4][5] have been done to find out relationships among network measurement algorithms.

In this research, the relationships among network measurement algorithms have been introduced for brain networks [6][7]. Besides discovering the relationships among these network measurement algorithms for each brain network, it is also shown how this study of relationships helps to understand brain networks combinedly by comparing the summary of statistical observations for each brain network through analyzing relationships among network measurement algorithms.

Sixteen several types of brain networks have been selected where six brain networks are human brain networks which have been collected as components from large human brain networks [6][7].These Sixteen brain networks are addressed as

1. bn-cat-mixed-species-brain-1
2. bn-fly-drosophila_medulla_1
3. component1-network-of-human-BNU-1-0025864-session-1-bg
4. component-2-network-of-human-BNU-1-0025864-session-1-bg,
5. component-3-network-of-human-BNU-1-0025864-session-1-bg
6. component-1-network-of-bn-human-BNU_1_0025864_session_2-bg
7. component-2-network-of-bn-human-BNU_1_0025864_session_2-bg
8. component-3-network-of-bn-human-BNU_1_0025864_session_2-bg
9. bn-macaque-rhesus_brain_1
10. bn-macaque-rhesus_brain_2
11. macaque-rhesus-cerebral-cortex-1
12. macaque-rhesus-interareal-cortical-network-2
13. bn-mouse-kasthuri_graph_v4
14. bn-mouse_brain_1
15. bn-mouse_visual-cortex_1,
16. bn-mouse_visual-cortex_2

## II. Methodology

Six network measurement algorithms which are Degree centrality[4], Pagerank Centrality[9], Betweenness Centrality[10], Closeness Centrality[11], Eigenvector

Centrality[12], Information Centrality[8][13][14] have been selected to apply on previous sixteen brain networks and the results of these sixteen brain networks after applying six network measurement algorithms are put into correlation method named Pearson Correlation Coefficient [15].

**Network Measurement Algorithms**

Let, G is a Graph with V vertices and E edges where $G = (V, E)$, $V = \{m_1, m_2, m_3, ..., m_n\}$ and $E = \{e_1, e_2, e_3, ..., e_n\}$

1. **Degree Centrality**[4]

$$Deg.\ Cen.\ (m_k) = \frac{deg(m_k)}{N-1}, m_k \in V$$

2. **Pagerank Centrality**[9]

$$Pag.\ Cen.\ (m_k) = \sum \frac{Pag.\ Cen.\ (m_i)}{deg(m_i)}, m_k \in V$$

3. **Betweenness Centrality**[10]

$$Bet.\ Cen.\ (m_k) = \frac{2 \sum_{j=1}^{N} \sum_{i=1}^{j-1} \frac{num\ of Sd_{ij}(m_k)}{num\ of Sd_{ij}}}{N^2 - 3N + 2}, i \neq j \neq k\ and\ \imath$$

4. **Closeness Centrality**[11]

$$Clo.\ Cen.\ (m_k) = \frac{N-1}{\sum_{i=1}^{N} Sd(m_i m_k)}, m_k \in V$$

5. **Eigenvector Centrality**[12]

$$Eig.\ Cen.\ (m_k) = \frac{1}{\lambda} \sum_{j=1}^{N} Connection_{kj} \times Eig.\ Cen.$$

6. **Information Centrality**[8][13][14]

$$Info.\ Cen.\ (m_s) = \frac{N}{\sum_{t \neq s, t=1}^{N} Throughput\ of\ m_s - Throughput\ of\ }$$

**Pearson Correlation Coefficient** [15]

$$r_{xy} = \frac{\sum_{i=1}^{N}(x_i - \bar{x})(y_i - \bar{y})}{\sqrt{\sum_{i=1}^{N}(x_i - \bar{x})^2} \sqrt{\sum_{i=1}^{N}(y_i - \bar{y})^2}}$$

where, x and y are results of two network measurement algorithms sequentially for a specific brain network

### III. EXPERIMENT

For experiment, Several Python libraries which are networkx [20] version 2.6.3, pandas [21] version 1.3.5, numpy [22] version 1.21.6, google.colab [23] version 0.0.1a2, matplotlib.pyplot [24], collections, io, tqdm [25] version 4.64.1 have been used in Google Colab platform [23]. Documentations done in Google Docs [26].

In this experiment, each brain network has been constructed from edges connection list. For the human brain networks, an algorithm similar to Breadth First Search(BFS) [16] is used to construct component networks from a large human brain network. For betweenness and eigenvector centrality algorithms, networks which have multigraphs have been considered into non-multigraphs and for information centrality, network with several components have been constructed into network with only one component by adding each node of small components into previous node of large component if and only if the number of nodes in large component in network >> the summation of number of all nodes in other components in network. After proper construction in several cases, all six algorithms have been applied to all sixteen brain networks. The result of each brain network after applying six network measurement algorithms can be represented into matrix which dimension is $6 \times total\ number\ of\ nodes\ in\ network$. By taking values of correspondent rows of considered two algorithms from each brain network's result that achieved previously, pearson correlation coefficients has been calculated only for those two algorithms applied into that specific brain network. At the end, correlation coefficient matrix with $6 \times 6$ dimension has been achieved for each brain network. In result section, only correlation coefficient matrix with $6 \times 6$ dimension has been shown.

### IV. RESULTS

In Tables, short forms have been used where
deg cen ← degree centrality
pag cen ← pagerank centrality
bet cen ← betweenness centrality
clo cen ← closeness centrality
eig cen ← eigenvector centrality
inf cen ← information centrality or,
info cen ← information centrality

**Result of network 1:**
Number of nodes = 65
Execution time needed for correlation matrix:
 00 minute 04 seconds

Correlation matrix for
bn-cat-mixed-species-brain-1:

|  | deg cen | pag cen | bet cen | clo cen | eig cen | inf cen |
|---|---|---|---|---|---|---|
| deg cen | 1.0 | 1.0 | 0.8 | 0.96 | 0.96 | 0.95 |
| pag cen | 1.0 | 1.0 | 0.83 | 0.95 | 0.94 | 0.94 |
| bet cen | 0.8 | 0.83 | 1.0 | 0.82 | 0.72 | 0.64 |
| clo cen | 0.96 | 0.95 | 0.82 | 1.0 | 0.96 | 0.93 |
| eig cen | 0.96 | 0.94 | 0.72 | 0.96 | 1.0 | 0.94 |
| inf cen | 0.95 | 0.94 | 0.64 | 0.93 | 0.94 | 1.0 |

Table 1:Correlation matrix for cat-mixed-species-brain-1

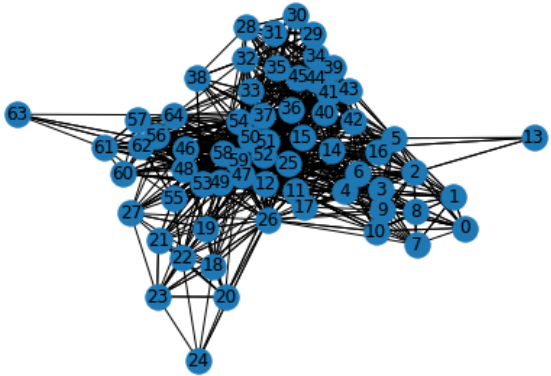

Figure 1: network visualization of bn-cat-mixed-species-brain-1

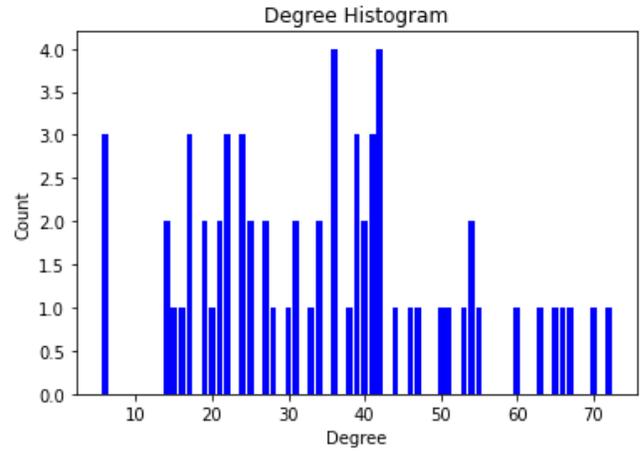

Figure 2: Degree Distribution Histogram of bn-cat-mixed-species-brain-1

**Result of network 2:**
Number of nodes=1781
Network type: multi graph

Execution time needed for correlation matrix while considering 50 nodes:
19 minutes 25 seconds
Correlation matrix for bn-fly-drosophila_medulla_1 with 50 nodes:

|  | deg cen | pag cen | bet cen | clo cen | eig cen | inf cen |
|---|---|---|---|---|---|---|
| deg cen | 1.0 | 0.99 | 0.79 | 0.24 | 0.81 | 0.47 |
| pag cen | 0.99 | 1.0 | 0.86 | 0.26 | 0.83 | 0.48 |
| bet cen | 0.79 | 0.86 | 1.0 | 0.31 | 0.81 | 0.42 |
| clo cen | 0.24 | 0.26 | 0.31 | 1.0 | 0.54 | 0.62 |
| eig cen | 0.81 | 0.83 | 0.81 | 0.54 | 1.0 | 0.49 |
| inf cen | 0.47 | 0.48 | 0.42 | 0.62 | 0.49 | 1.0 |

Table 2:Correlation matrix of bn-fly-drosophila_medulla_1 for 50 nodes

Execution time needed for correlation matrix while considering 100 nodes:
38 minutes 42 seconds

Correlation matrix of bn-fly-drosophila_medulla_1 with 100 nodes:

|       | deg cen | pag cen | bet cen | clo cen | eig cen | inf cen |
|-------|---------|---------|---------|---------|---------|---------|
| deg cen | 1.0   | 1.0     | 0.88    | 0.4     | 0.87    | 0.47    |
| pag cen | 1.0   | 1.0     | 0.91    | 0.41    | 0.88    | 0.49    |
| bet cen | 0.88  | 0.91    | 1.0     | 0.42    | 0.83    | 0.44    |
| clo cen | 0.4   | 0.41    | 0.42    | 1.0     | 0.62    | 0.78    |
| eig cen | 0.87  | 0.88    | 0.83    | 0.62    | 1.0     | 0.58    |
| inf cen | 0.47  | 0.49    | 0.44    | 0.78    | 0.58    | 1.0     |

Table 3: Correlation matrix of bn-fly-drosophila_medulla_1 for 100 nodes

Execution time needed for correlation matrix while considering 500 nodes:
4 hours 39 minutes 17 seconds
Correlation matrix for bn-fly-drosophila_medulla_1 with 500 nodes:

|       | deg cen | pag cen | bet cen | clo cen | eig cen | inf cen |
|-------|---------|---------|---------|---------|---------|---------|
| deg cen | 1.0   | 0.99    | 0.81    | 0.32    | 0.81    | 0.44    |
| pag cen | 0.99  | 1.0     | 0.86    | 0.33    | 0.82    | 0.46    |
| bet cen | 0.81  | 0.86    | 1.0     | 0.35    | 0.79    | 0.44    |
| clo cen | 0.32  | 0.33    | 0.35    | 1.0     | 0.58    | 0.81    |
| eig cen | 0.81  | 0.82    | 0.79    | 0.58    | 1.0     | 0.63    |
| inf cen | 0.44  | 0.46    | 0.44    | 0.81    | 0.63    | 1.0     |

Table 4: Correlation matrix for bn-fly-drosophila_medulla_1 with 500 nodes

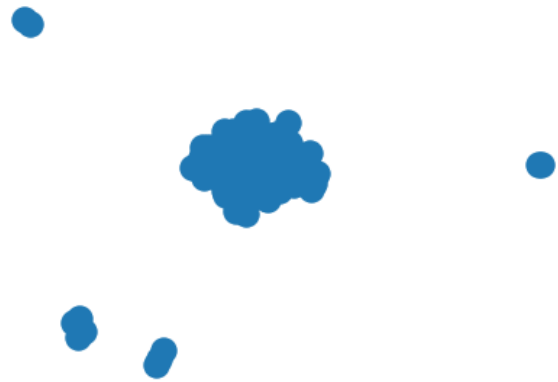

Figure 3: network visualization of bn-fly-drosophila_medulla_1

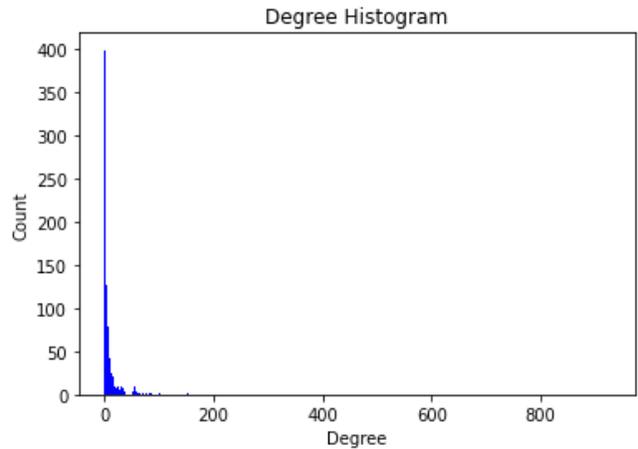

Figure 4: Degree Distribution Histogram of bn-fly-drosophila_medulla_1

**Result of network 3:**

Component 1 network of human-BNU-1-0025864-session-1-bg:
number of nodes: 58
execution time needed for correlation matrix:
00 minute 05 seconds

Correlation matrix for component 1 in human-BNU-1-0025864-session-1-bg:

|       | deg cen | pag cen | bet cen | clo cen | eig cen | inf cen |
|-------|---------|---------|---------|---------|---------|---------|
| deg cen | 1.0   | 0.96    | 0.84    | 0.72    | 0.79    | 0.85    |

|  | pag cen | bet cen | clo cen | eig cen | inf cen |
|---|---|---|---|---|---|
| pag cen | 0.96 | 1.0 | 0.89 | 0.59 | 0.66 | 0.73 |
| bet cen | 0.84 | 0.89 | 1.0 | 0.55 | 0.58 | 0.61 |
| clo cen | 0.72 | 0.59 | 0.55 | 1.0 | 0.88 | 0.9 |
| eig cen | 0.79 | 0.66 | 0.58 | 0.88 | 1.0 | 0.79 |
| inf cen | 0.85 | 0.73 | 0.61 | 0.9 | 0.79 | 1.0 |

Table 5: Correlation matrix for component 1 in human-BNU-1-0025864-session-1-bg

Figure 5: component 1 network visualization of human-BNU-1-0025864-session-1-bg

Figure 6: Degree Distribution Histogram of component 1 network in human-BNU-1-0025864-session-1-bg

**Result of network 4:**

Component 2 network of human-BNU-1-0025864-session-1-bg:

number of nodes: 359
execution time needed for correlation matrix: 15 minutes 59 seconds

Correlation matrix for component 2 in human-BNU-1-0025864-session-1-bg:

|  | deg cen | pag cen | bet cen | clo cen | eig cen | inf cen |
|---|---|---|---|---|---|---|
| deg cen | 1.0 | 0.89 | 0.49 | 0.47 | 0.69 | 0.67 |
| pag cen | 0.89 | 1.0 | 0.54 | 0.23 | 0.42 | 0.47 |
| bet cen | 0.49 | 0.54 | 1.0 | 0.23 | 0.2 | 0.24 |
| clo cen | 0.47 | 0.23 | 0.23 | 1.0 | 0.54 | 0.85 |
| eig cen | 0.69 | 0.42 | 0.2 | 0.54 | 1.0 | 0.55 |
| inf cen | 0.67 | 0.47 | 0.24 | 0.85 | 0.55 | 1.0 |

Table 6: Correlation matrix for component 2 in human-BNU-1-0025864-session-1-bg

Figure 7: component 2 network visualization of human-BNU-1-0025864-session-1-bg

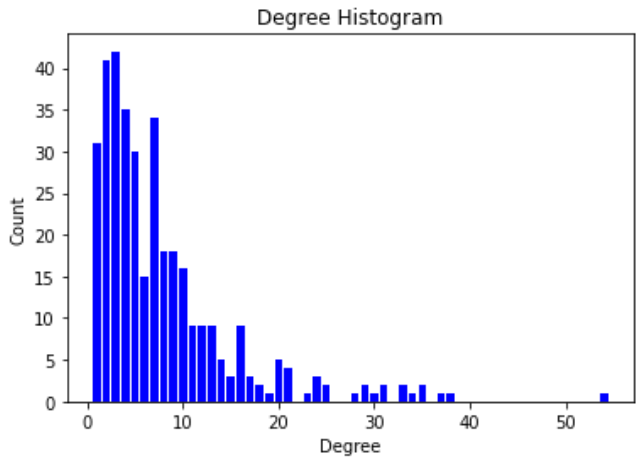

Figure 8: Degree Distribution Histogram of component 2 network in human-BNU-1-0025864-session-1-bg

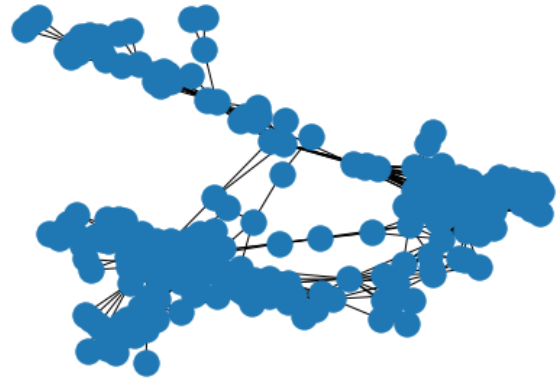

Figure 9: component 3 network visualization of human-BNU-1-0025864-session-1-bg

**Result of network 5:**
Component 3 network of
human-BNU-1-0025864-session-1-bg:

number of nodes: 269
execution time needed for correlation matrix:
07 minutes 57 seconds

Correlation matrix of component 3 in
human-BNU-1-0025864-session-1-bg:

|  | deg cen | pag cen | bet cen | clo cen | eig cen | inf cen |
|---|---|---|---|---|---|---|
| deg cen | 1.0 | 0.83 | 0.34 | 0.63 | 0.7 | 0.82 |
| pag cen | 0.83 | 1.0 | 0.5 | 0.41 | 0.33 | 0.58 |
| bet cen | 0.34 | 0.5 | 1.0 | 0.4 | 0.04 | 0.26 |
| clo cen | 0.63 | 0.41 | 0.4 | 1.0 | 0.38 | 0.86 |
| eig cen | 0.7 | 0.33 | 0.04 | 0.38 | 1.0 | 0.54 |
| inf cen | 0.82 | 0.58 | 0.26 | 0.86 | 0.54 | 1.0 |

Table 7: Correlation matrix for component 3 in
human-BNU-1-0025864-session-1-bg

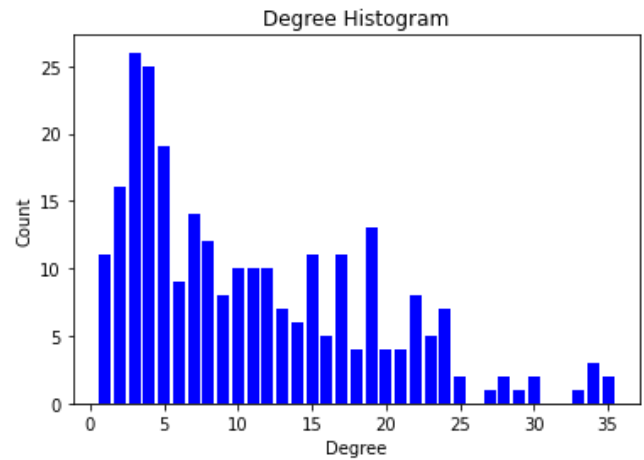

Figure 10: Degree Distribution Histogram of component 3 network in human-BNU-1-0025864-session-1-bg

**Result of network 6:**

Number of nodes: 242
Network type: undirected graph
Execution time needed for correlation matrix:
03 minutes 11 seconds

Correlation matrix for
bn-macaque-rhesus_brain_1:

|  | deg cen | pag cen | bet cen | clo cen | eig cen | info cen |
|---|---|---|---|---|---|---|
| deg cen | 1.0 | 1.0 | 0.85 | 0.88 | 0.94 | 0.83 |
| pag cen | 1.0 | 1.0 | 0.87 | 0.87 | 0.91 | 0.82 |
| bet cen | 0.85 | 0.87 | 1.0 | 0.66 | 0.75 | 0.52 |
| clo cen | 0.88 | 0.87 | 0.66 | 1.0 | 0.92 | 0.9 |
| eig cen | 0.94 | 0.91 | 0.75 | 0.92 | 1.0 | 0.81 |
| info cen | 0.83 | 0.82 | 0.52 | 0.9 | 0.81 | 1.0 |

Table 8:Correlation matrix for bn-macaque-rhesus_brain_1

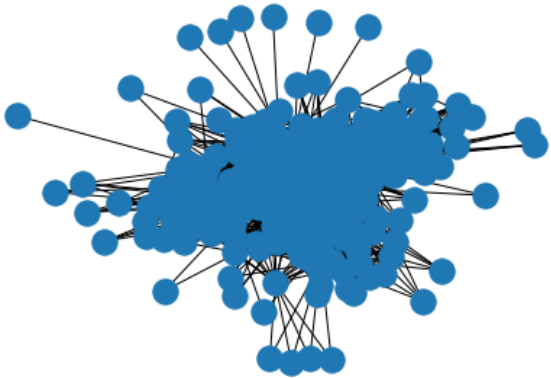

Figure 11: network visualization of bn-macaque-rhesus_brain_1

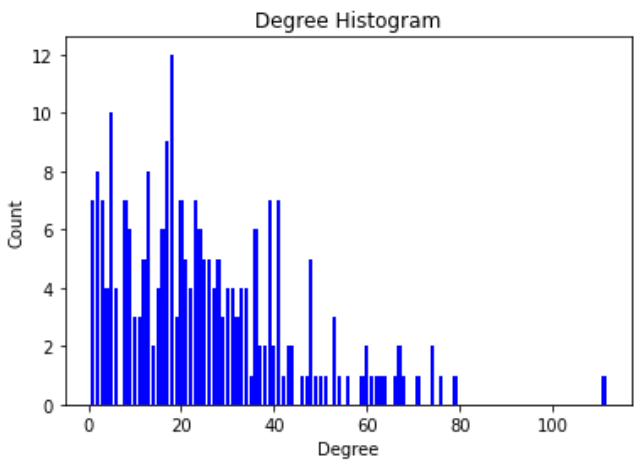

Figure 12: Degree Distribution Histogram of network bn-macaque-rhesus_brain_1

**Result of network 7:**

number of nodes: 91
network type: undirected graph
execution time needed for correlation matrix:
00 minute 09 seconds

Correlation matrix for bn-macaque-rhesus_brain_2:

|   | deg cen | pag cen | bet cen | clo cen | eig cen | info cen |
|---|---|---|---|---|---|---|
| deg cen | 1.0 | 1.0 | 0.89 | 0.99 | 0.96 | 0.81 |
| pag cen | 1.0 | 1.0 | 0.92 | 0.99 | 0.95 | 0.79 |
| bet cen | 0.89 | 0.92 | 1.0 | 0.94 | 0.78 | 0.58 |
| clo cen | 0.99 | 0.99 | 0.94 | 1.0 | 0.93 | 0.78 |
| eig cen | 0.96 | 0.95 | 0.78 | 0.93 | 1.0 | 0.94 |
| info cen | 0.81 | 0.79 | 0.58 | 0.78 | 0.94 | 1.0 |

Table 9:Correlation matrix for bn-macaque-rhesus_brain_2

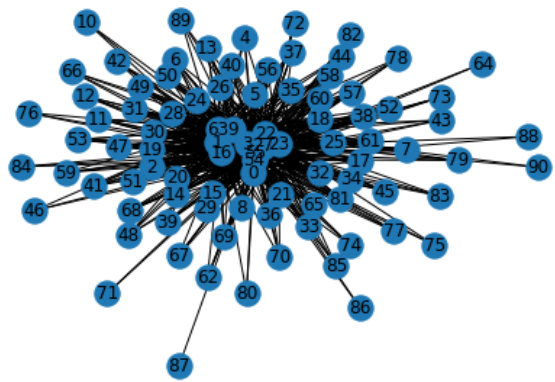

Figure 13: network visualization of bn-macaque-rhesus_brain_2

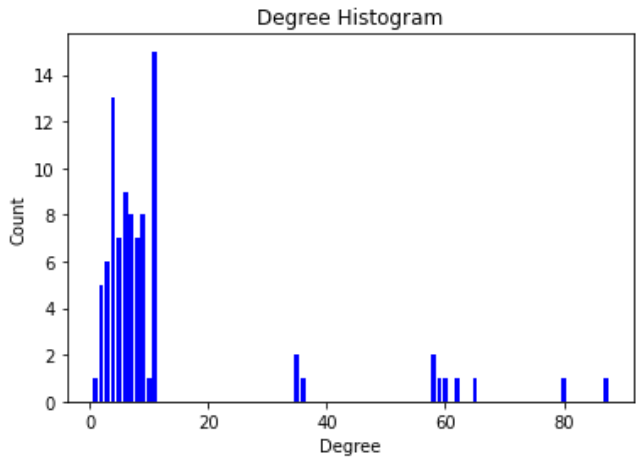

Figure 14: Degree Distribution Histogram of network bn-macaque-rhesus_brain_2

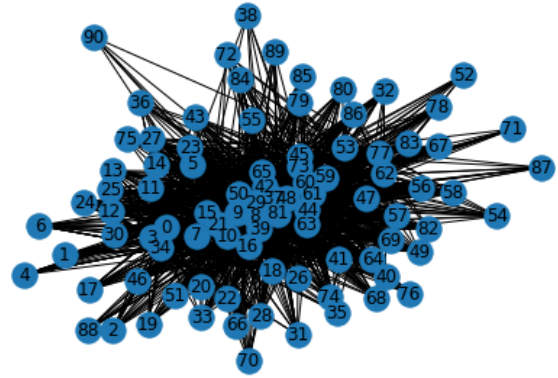

Figure 15: network visualization of macaque-rhesus-cerebral-cortex-1

**Result of network 8:**

Number of nodes: 91
Network type: undirected Multigraph
Execution time needed for correlation matrix:
00 minute 18 seconds

Correlation matrix for macaque-rhesus-cerebral-cortex-1:

|  | deg cen | pag cen | bet cen | clo cen | eig cen | info cen |
|---|---|---|---|---|---|---|
| deg cen | 1.0 | 1.0 | 0.77 | 0.82 | 0.81 | 0.88 |
| pag cen | 1.0 | 1.0 | 0.78 | 0.83 | 0.81 | 0.87 |
| bet cen | 0.77 | 0.78 | 1.0 | 0.97 | 0.89 | 0.77 |
| clo cen | 0.82 | 0.83 | 0.97 | 1.0 | 0.97 | 0.88 |
| eig cen | 0.81 | 0.81 | 0.89 | 0.97 | 1.0 | 0.94 |
| info cen | 0.88 | 0.87 | 0.77 | 0.88 | 0.94 | 1.0 |

Table 10: Correlation matrix for macaque-rhesus-cerebral-cortex-1

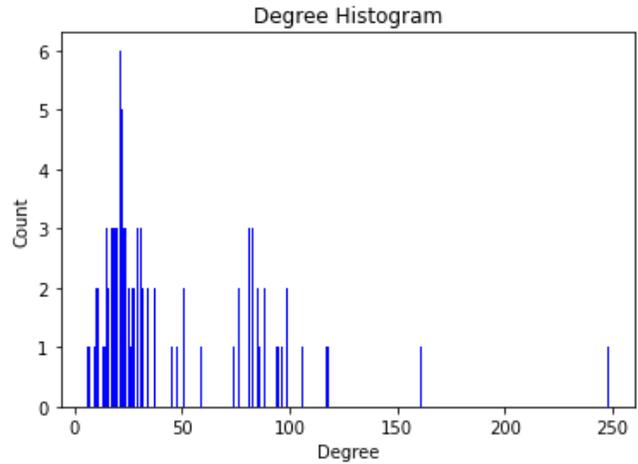

Figure 16: Degree Distribution Histogram of network macaque-rhesus-cerebral-cortex-1

**Result of network 9:**

number of nodes: 93
network type: undirected Multigraph
execution time needed for correlation matrix:
00 minute 26 seconds

Correlation matrix for macaque-rhesus-interareal-cortical-network-2:

|  | deg cen | pag cen | bet cen | clo cen | eige cen | info cen |
|---|---|---|---|---|---|---|
| deg cen | 1.0 | 1.0 | 1.0 | 1.0 | 1.0 | 1.0 |
| pag cen | 1.0 | 1.0 | 1.0 | 1.0 | 1.0 | 1.0 |
| bet cen  | 1.0 | 1.0 | 1.0 | 1.0 | 1.0 | 1.0 |
| clo cen  | 1.0 | 1.0 | 1.0 | 1.0 | 1.0 | 1.0 |
| eig cen  | 1.0 | 1.0 | 1.0 | 1.0 | 1.0 | 1.0 |
| info cen | 1.0 | 1.0 | 1.0 | 1.0 | 1.0 | 1.0 |

Table 11: Correlation matrix for macaque-rhesus-interareal-cortical-network-2

**Result of network 10:**

number of nodes: 1029
network type: undirected Multigraph
execution time needed for correlation matrix:
02 hours 00 minutes 47 seconds

Correlation matrix for bn-mouse-kasthuri_graph_v4:

|         | deg cen | pag cen | bet cen | clo cen | eige cen | info cen |
|---------|---------|---------|---------|---------|----------|----------|
| deg cen | 1.0     | 0.99    | 0.97    | 0.27    | 0.68     | 0.47     |
| pag cen | 0.99    | 1.0     | 0.96    | 0.2     | 0.63     | 0.43     |
| bet cen | 0.97    | 0.96    | 1.0     | 0.26    | 0.74     | 0.4      |
| clo cen | 0.27    | 0.2     | 0.26    | 1.0     | 0.46     | 0.7      |
| eig cen | 0.68    | 0.63    | 0.74    | 0.46    | 1.0      | 0.43     |
| info cen| 0.47    | 0.43    | 0.4     | 0.7     | 0.43     | 1.0      |

Table 12: Correlation matrix of bn-mouse-kasthuri_graph_v4

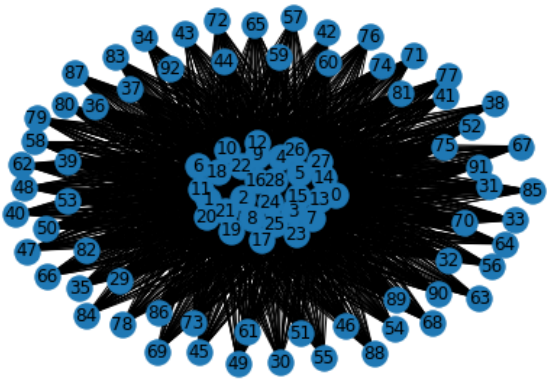

Figure 17: network visualization of macaque-rhesus-interareal-cortical-network-2

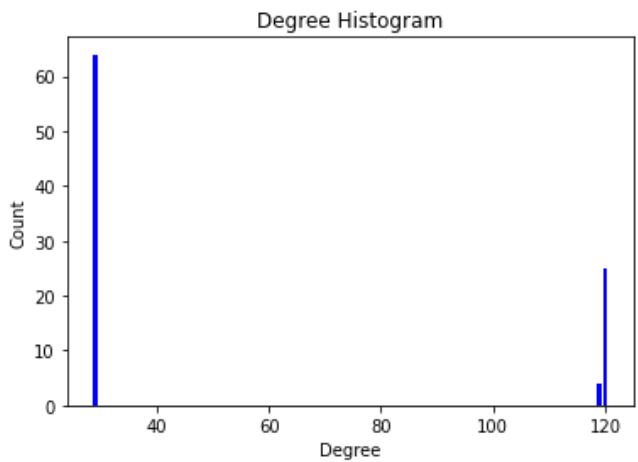

Figure 18: Degree Distribution Histogram of network macaque-rhesus-interareal-cortical-network-2

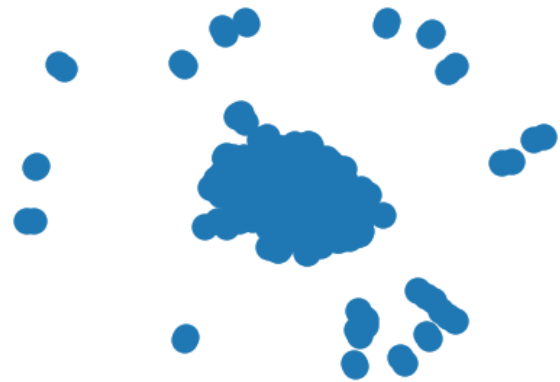

Figure 19: network visualization for bn-mouse-kasthuri_graph_v4 multigraph with 20 components

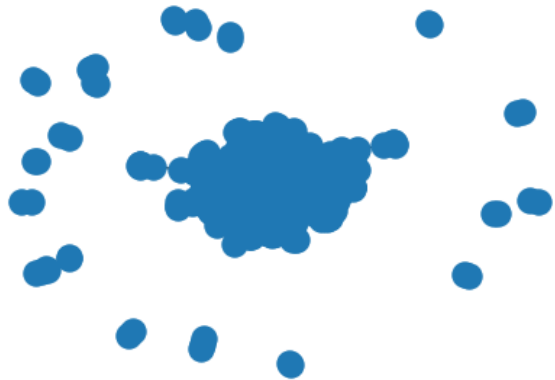

Figure 20: network visualization for
bn-mouse-kasthuri_graph_v4 ordinary graph with 20
components
(needed for betweenness centrality and eigenvector
centrality)

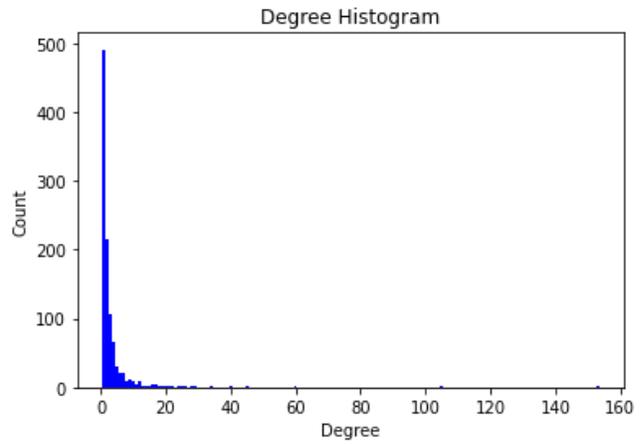

Figure 22: Degree Distribution Histogram of network
bn-mouse-kasthuri_graph_v4 multigraph with 20
components

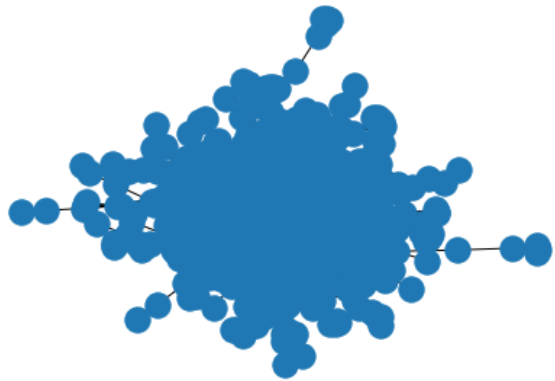

Figure 21: network visualization of constructed
bn-mouse-kasthuri_graph_v4 multigraph with the largest
component
(needed for information centrality)

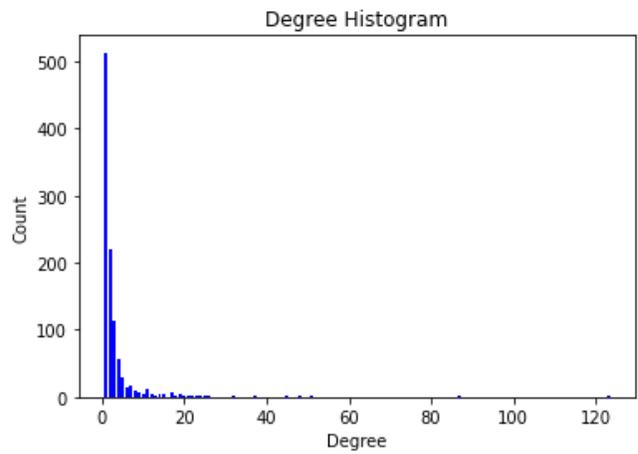

Figure 23: Degree Distribution Histogram of network
bn-mouse-kasthuri_graph_v4 ordinary graph with 20
components

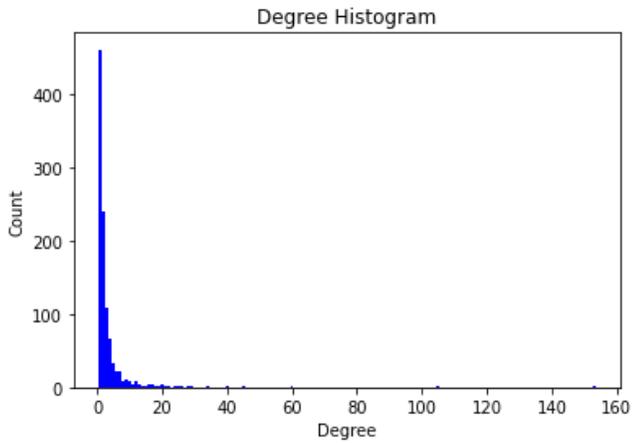

Figure 24: Degree Distribution Histogram of constructed network bn-mouse-kasthuri_graph_v4 multigraph with the largest component

Table13:Correlation matrix for bn-mouse_brain_1

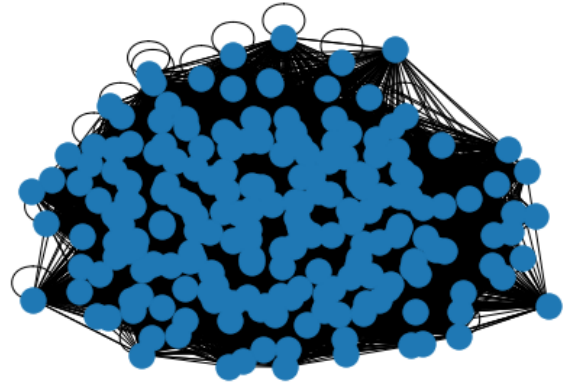

Figure 25: network visualization of bn-mouse_brain_1 multigraph

**Result of network 11:**

number of nodes: 213
network type: undirected Multigraph
execution time needed for correlation matrix:
09 minutes 03 seconds

Correlation matrix for bn-mouse_brain_1:

|  | deg cen | pag cen | bet cen | clo cen | eige cen | info cen |
| --- | --- | --- | --- | --- | --- | --- |
| deg cen | 1.0 | 1.0 | 0.95 | 0.97 | 0.97 | 0.99 |
| pag cen | 1.0 | 1.0 | 0.95 | 0.97 | 0.97 | 0.99 |
| bet cen | 0.95 | 0.95 | 1.0 | 0.98 | 0.95 | 0.92 |
| clo cen | 0.97 | 0.97 | 0.98 | 1.0 | 0.99 | 0.96 |
| eig cen | 0.97 | 0.97 | 0.95 | 0.99 | 1.0 | 0.97 |
| info cen | 0.99 | 0.99 | 0.92 | 0.96 | 0.97 | 1.0 |

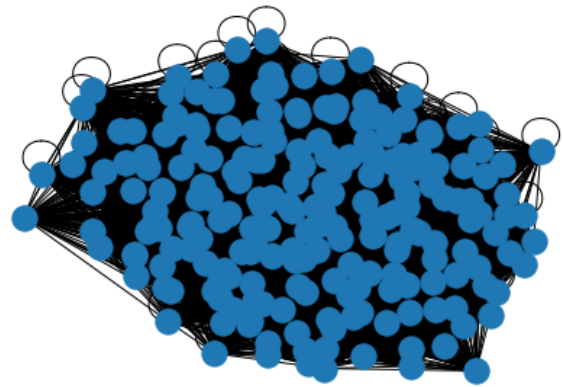

Figure 26: network visualization for bn-mouse_brain_1 graph

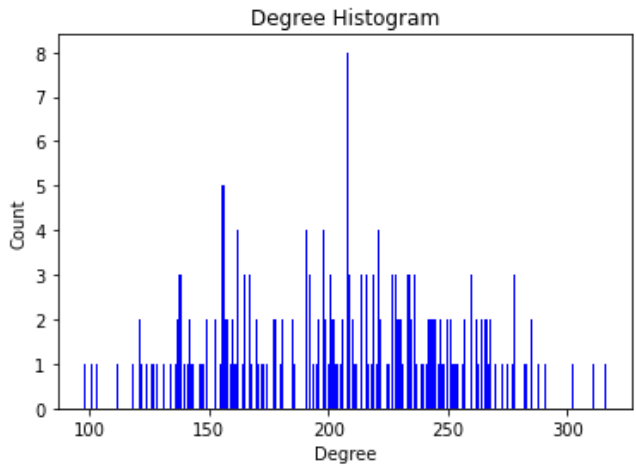

Figure 27: Degree Distribution Histogram of network bn-mouse_brain_1 multigraph

|        | deg cen | pag cen | bet cen | clo cen | eig cen | info cen |
|--------|---------|---------|---------|---------|---------|----------|
| deg cen | 1.0 | 0.99 | 0.9 | 0.76 | 0.84 | 0.86 |
| pag cen | 0.99 | 1.0 | 0.91 | 0.72 | 0.79 | 0.82 |
| bet cen | 0.9 | 0.91 | 1.0 | 0.84 | 0.82 | 0.79 |
| clo cen | 0.76 | 0.72 | 0.84 | 1.0 | 0.91 | 0.92 |
| eig cen | 0.84 | 0.79 | 0.82 | 0.91 | 1.0 | 0.88 |
| info cen | 0.86 | 0.82 | 0.79 | 0.92 | 0.88 | 1.0 |

Table 14: Correlation matrix for bn-mouse_visual-cortex_1

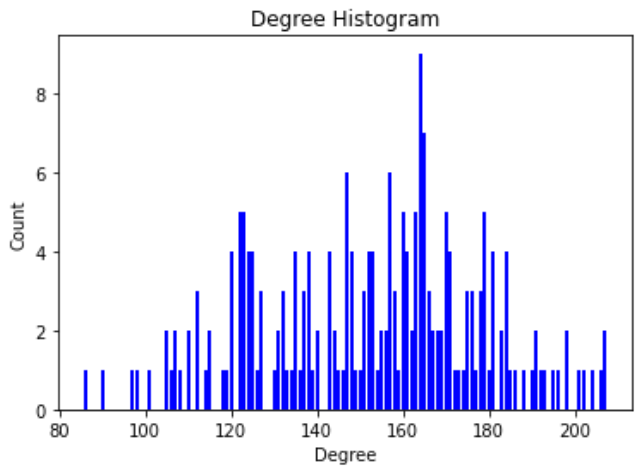

Figure 28: Degree Distribution Histogram of network bn-mouse_brain_1 graph

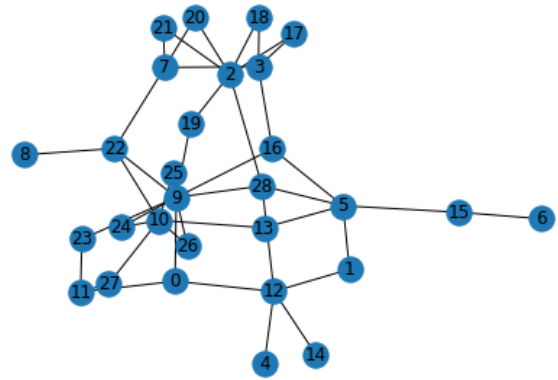

Figure 29: network visualization of bn-mouse_visual-cortex_1

**Result of network 12:**

number of nodes: 29
network type: undirected graph
execution time needed for correlation matrix: <<00 minute 01 second

Correlation matrix for bn-mouse_visual-cortex_1:

|  | deg cen | pag cen | bet cen | clo cen | eig cen | info cen |
|--|---------|---------|---------|---------|---------|----------|

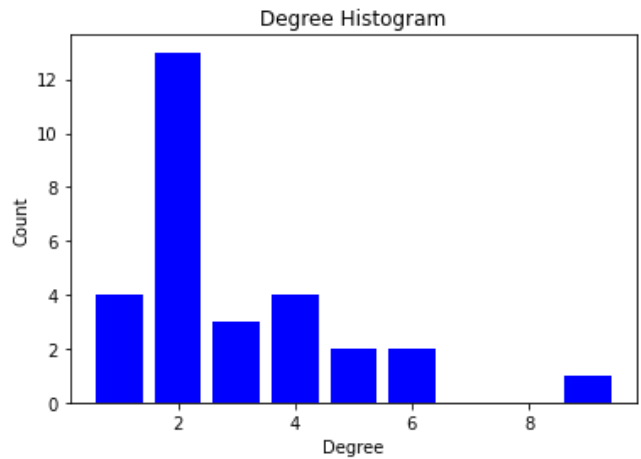

Figure 30: Degree Distribution Histogram of network bn-mouse_visual-cortex_1

**Result of network 13:**

number of nodes: 193
network type: undirected multigraph
execution time needed for correlation matrix:
<<00 minutes 38 seconds

Correlation matrix for bn-mouse_visual-cortex_2:

|  | deg cen | pag cen | bet cen | clo cen | eig cen | info cen |
|---|---|---|---|---|---|---|
| deg cen | 1.0 | 1.0 | 0.9 | 0.4 | 0.75 | 0.8 |
| pag cen | 1.0 | 1.0 | 0.89 | 0.39 | 0.73 | 0.79 |
| bet cen | 0.9 | 0.89 | 1.0 | 0.51 | 0.75 | 0.84 |
| clo cen | 0.4 | 0.39 | 0.51 | 1.0 | 0.65 | 0.62 |
| eig cen | 0.75 | 0.73 | 0.75 | 0.65 | 1.0 | 0.78 |
| info cen | 0.8 | 0.79 | 0.84 | 0.62 | 0.78 | 1.0 |

Table 15: Correlation matrix for bn-mouse_visual-cortex_2

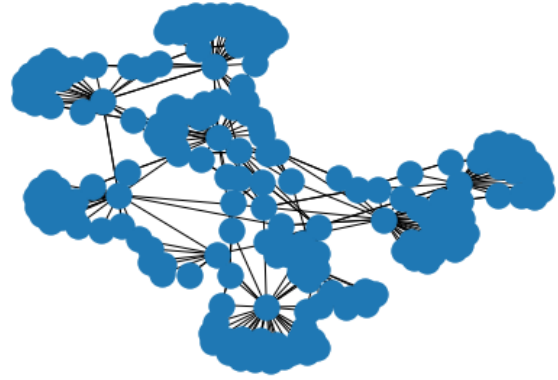

Figure 32: network visualization of bn-mouse_visual-cortex_2 with only largest component

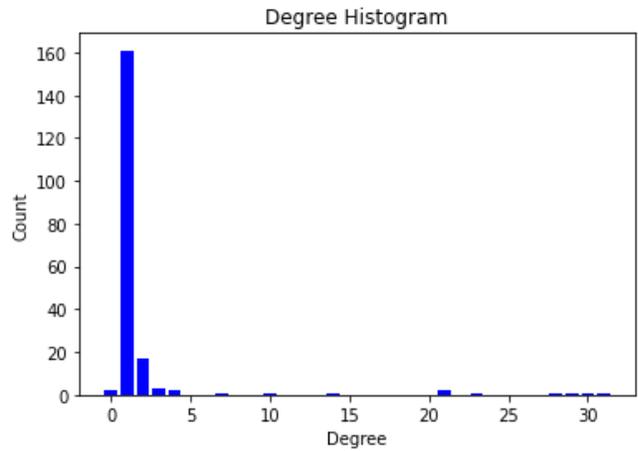

Figure 33: Degree Distribution Histogram of network bn-mouse_visual-cortex_2

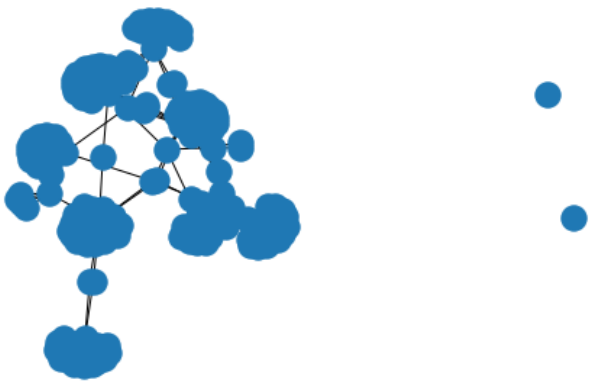

Figure 31: network visualization of bn-mouse_visual-cortex_2

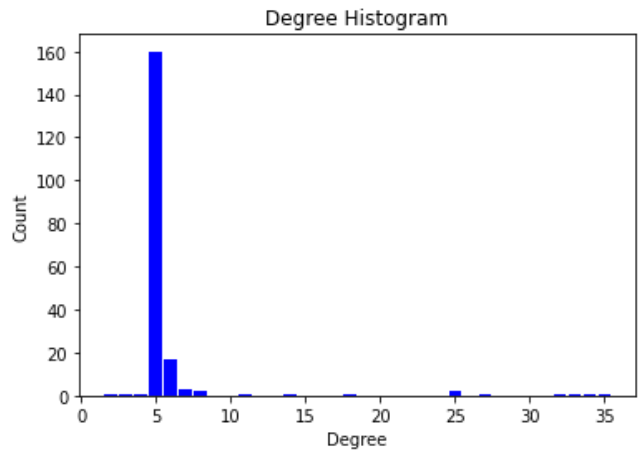

Figure 34: Degree Distribution Histogram of network bn-mouse_visual-cortex_2 with only largest component

Component 1 network of
bn-human-BNU_1_0025864_session_2-bg:

Number of nodes: 30
Execution time needed for correlation matrix:
00 minute 01 second

Correlation matrix for component 1 network of
bn-human-BNU_1_0025864_session_2-bg:

|         | deg cen | pag cen | bet cen | clo cen | eig cen | info cen |
|---------|---------|---------|---------|---------|---------|----------|
| deg cen | 1.0     | 0.99    | 0.84    | 0.93    | 0.94    | 0.89     |
| pag cen | 0.99    | 1.0     | 0.89    | 0.91    | 0.9     | 0.85     |
| bet cen | 0.84    | 0.89    | 1.0     | 0.72    | 0.67    | 0.57     |
| clo cen | 0.93    | 0.91    | 0.72    | 1.0     | 0.95    | 0.91     |
| eig cen | 0.94    | 0.9     | 0.67    | 0.95    | 1.0     | 0.95     |
| info cen| 0.89    | 0.85    | 0.57    | 0.91    | 0.95    | 1.0      |

Table 16:Correlation matrix for component 1 network of
bn-human-BNU_1_0025864_session_2-bg

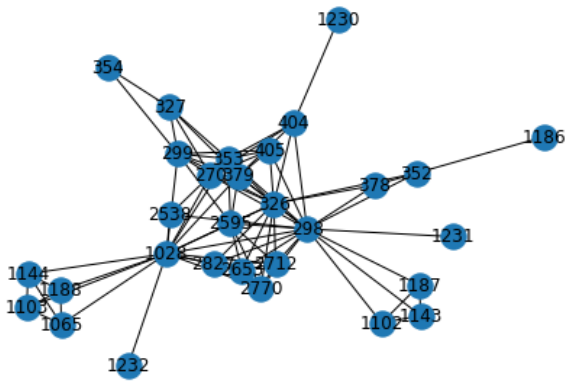

Figure 35: network visualization of component 1 of
bn-human-BNU_1_0025864_session_2-bg

**Result of network 14:**

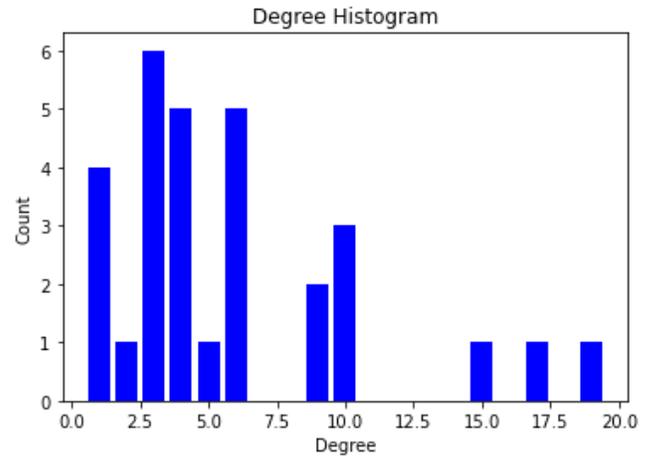

Figure 36: Degree Distribution Histogram of component 1
network in bn-human-BNU_1_0025864_ session_2-bg

**Result of network 15:**
Component 2 network of
bn-human-BNU_1_0025864_session_2-bg:
Number of nodes: 60
Execution time needed for correlation matrix:
00 minute 05 seconds

Correlation matrix for component 2 network of
bn-human-BNU_1_0025864_session_2-bg:

|         | deg cen | pag cen | bet cen | clo cen | eig cen | info cen |
|---------|---------|---------|---------|---------|---------|----------|
| deg cen | 1.0     | 0.85    | 0.08    | 0.77    | 0.93    | 0.82     |
| pag cen | 0.85    | 1.0     | 0.33    | 0.5     | 0.68    | 0.5      |
| bet cen | 0.08    | 0.33    | 1.0     | 0.21    | -0.06   | 0.01     |
| clo cen | 0.77    | 0.5     | 0.21    | 1.0     | 0.67    | 0.94     |
| eig cen | 0.93    | 0.68    | -0.06   | 0.67    | 1.0     | 0.74     |
| info cen| 0.82    | 0.5     | 0.01    | 0.94    | 0.74    | 1.0      |

Table 17:Correlation matrix for component 2 network of
bn-human-BNU_1_0025864_session_2-bg

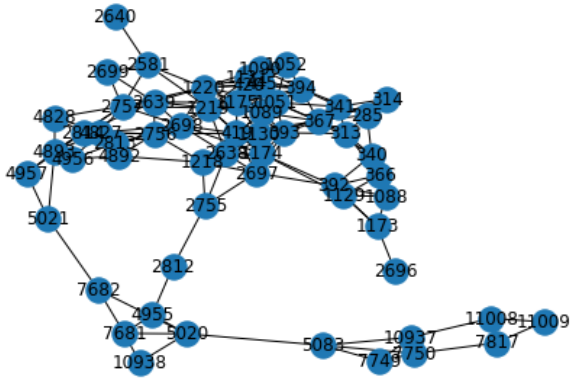

Figure 37: network visualization of component 2 of bn-human-BNU_1_0025864_session_2-bg

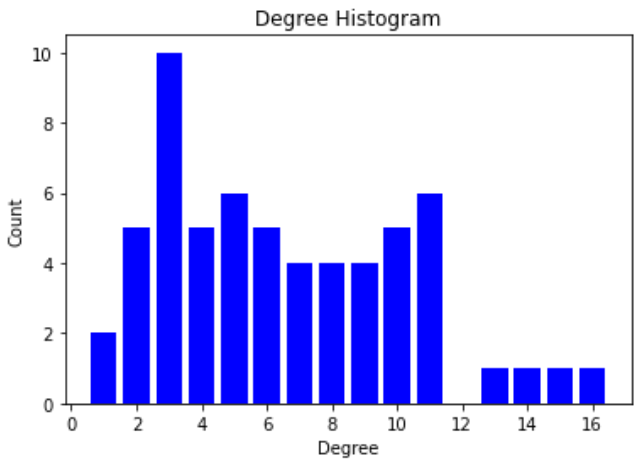

Figure 38: Degree Distribution Histogram of component 2 network in bn-human-BNU_1_0025864_ session_2-bg

**Result of network 16:**
Component 3 network of
bn-human-BNU_1_0025864_session_2-bg:
Number of nodes: 253
Execution time needed for correlation matrix:
05 minutes 48 seconds

Correlation matrix for component 3 network of bn-human-BNU_1_0025864_session_2-bg:

|        | deg cen | pag cen | bet cen | clo cen | eig cen | info cen |
|--------|---------|---------|---------|---------|---------|----------|
| deg cen | 1.0 | 0.89 | 0.57 | 0.61 | 0.83 | 0.65 |
| pag cen | 0.89 | 1.0 | 0.62 | 0.36 | 0.53 | 0.45 |
| bet cen | 0.57 | 0.62 | 1.0 | 0.31 | 0.31 | 0.26 |
| clo cen | 0.61 | 0.36 | 0.31 | 1.0 | 0.72 | 0.91 |
| eig cen | 0.83 | 0.53 | 0.31 | 0.72 | 1.0 | 0.67 |
| info cen | 0.65 | 0.45 | 0.26 | 0.91 | 0.67 | 1.0 |

Table 18: Correlation matrix for component 3 network of bn-human-BNU_1_0025864_session_2-bg

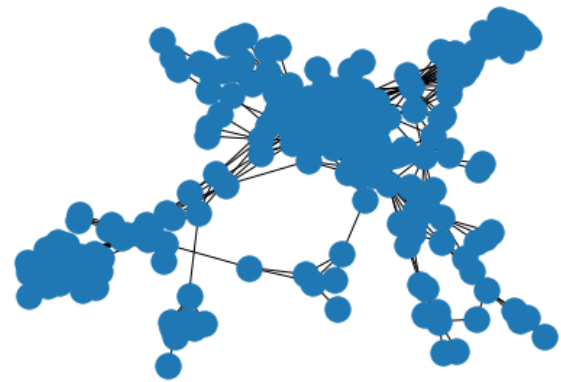

Figure 39: network visualization of component 3 of bn-human-BNU_1_0025864_session_2-bg

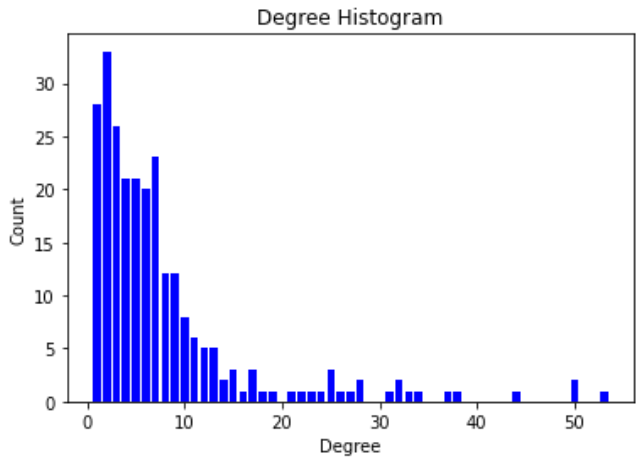

Figure 40: Degree Distribution Histogram of component 3 network in bn-human-BNU_1_0025864_ session_2-bg

Statistical Comparison Chart among brain networks based on number of nodes, degree distributions, high correlation coefficients among applied network measurement algorithms and the summary of statistical observations by analyzing correlations among network measurement algorithms for each brain network:

| Name of the brain Network | Number of Nodes | Degree Distribution | High Correlation Coefficients among network measurement algorithms | Number of High Correlation Coefficients among Network Measurement Algorithms | Summary of Statistical Observations by Analyzing Correlations among Network Measurement Algorithms |
|---|---|---|---|---|---|
| bn-cat-mixed-species-brain-1 | 65 | Slightly Normal Distribution | deg cen vs pag cen, deg cen vs bet cen, deg cen vs clo cen, deg cen vs eig cen, deg cen vs inf cen, pag cen vs bet cen, pag cen vs clo cen, pag cen vs eig cen, pag cen vs inf cen, bet cen vs clo cen, clo cen vs eig cen, clo cen vs inf cen, eig cen vs inf cen | 13 | Nodes with high degree are mostly important nodes which are inclined to have less path difference form most of the nodes, having high tendency to be in the shortest path between any two nodes, having tendency to connect with important nodes and having more information in network |
| bn-fly-drosophila_medulla_1 | 1781 | Power Law Distribution | deg cen vs pag cen, deg cen vs bet cen, | 7 | Nodes with high degree are mostly |

| | | | deg cen vs eig cen, pag cen vs bet cen, pag cen vs eig cen, bet cen vs eig cen, clo cen vs info cen | | important nodes, having a high tendency to be in the shortest path between any two nodes and having a tendency to connect with important nodes, nodes which have less path differences from most other nodes are inclined to have more information. |
|---|---|---|---|---|---|
| Component 1 network of human-BNU-1-0025864-session-1-bg | 58 | Slightly Normal Distribution, Slightly Power Law Distribution | deg cen vs pag cen, deg cen vs bet cen, deg cen vs eig cen, deg cen vs inf cen, pag cen vs bet cen, clo cen vs eig cen, clo cen vs inf cen, eig cen vs inf cen | 8 | Nodes with high degree are mostly important nodes, having high tendency to be in the shortest path between any two nodes, having tendency to connect with important nodes and having more information in network, nodes which have less path differences from most other nodes are inclined to have more information and having tendency to connect with important nodes. |
| Component 2 network of human-BNU-1-0025864-session-1-bg | 359 | Power Law Distribution | deg cen vs pag cen, clo cen vs info cen | 2 | Nodes with high degree are mostly important nodes, nodes which have less path differences from most other nodes are having more information |
| Component 3 network of human-BNU-1-0025864-session-1-bg | 269 | Power Law Distribution | deg cen vs pag cen, deg cen vs info cen, clo cen vs info cen | 3 | Nodes with high degree are mostly important nodes, having more information, nodes |
| | | | | | which have less path differences from most other nodes are having more information |
| bn-macaque-rhesus_brain_1 | 242 | Slightly Power Law Distribution | deg cen vs pag cen, deg cen vs bet cen, deg cen vs clo cen, deg cen vs eig cen, deg cen vs info cen, pag cen vs bet cen, pag cen vs clo cen, pag cen vs eig cen, pag cen vs info cen, clo cen vs eig cen, clo cen vs info cen, eig cen vs info cen | 11 | Nodes with high degree are mostly important nodes which are inclined to have less path difference form most of the nodes, having high tendency to be in the shortest path between any two nodes, having tendency to connect with important nodes and having more information in network |
| bn-macaque-rhesus_brain_2 | 91 | Slightly Power Law Distribution | deg cen vs pag cen, deg cen vs bet cen, deg cen vs clo cen, deg cen vs eig cen, deg cen vs info cen, pag cen vs bet cen, pag cen vs clo cen, pag cen vs eig cen, pag cen vs info cen, bet cen vs clo cen, bet cen vs eig cen, clo cen vs eig cen, clo cen vs info cen, eig cen vs info cen | 14 | Nodes with high degree are mostly important nodes which are inclined to have less path difference form most of the nodes, having high tendency to be in the shortest path between any two nodes, having tendency to connect with important nodes and having more information in network |
| macaque-rhesus-cerebral- cortex-1 | 91 | Slightly Power Law Distribution | deg cen vs pag cen, deg cen vs clo cen, deg cen vs eig cen, deg cen vs info cen, pag cen vs clo cen, pag cen vs eig cen, pag cen vs info cen, bet cen vs clo cen, bet cen vs eig cen, clo cen vs eig cen, clo cen vs info cen, eig cen vs info cen | 12 | Nodes with high degree are mostly important nodes which are inclined to have less path difference form most of the nodes, having tendency to connect with important nodes and having more information in |

| | | | | | network |
|---|---|---|---|---|---|
| macaque-rhesus-interareal-cortical-network-2 | 93 | Slightly Power Law Distribution | all correlation coefficients are high | 15 | Nodes with high degree are mostly important nodes which are inclined to have less path difference form most of the nodes, having high tendency to be in the shortest path between any two nodes, having tendency to connect with important nodes and having more information in network |
| bn-mouse-kasthuri_graph_v4 | 1029 | Power Law Distribution | deg cen vs pag cen, deg cen vs bet cen, pag cen vs bet cen | 3 | Nodes with high degree are mostly important nodes and having high tendency to be in the shortest path between any two nodes |
| bn-mouse_brain_1 | 213 | Normal Distribution | all correlation coefficients are high | 15 | Nodes with high degree are mostly important nodes which are inclined to have less path difference form most of the nodes, having high tendency to be in the shortest path between any two nodes, having tendency to connect with important nodes and having more information |
| bn-mouse_visual-cortex_1 | 29 | Slightly Normal Distribution, Slightly Power Law Distribution | deg cen vs pag cen, deg cen vs bet cen, deg cen vs eig cen, deg cen vs info cen, pag cen vs bet cen, pag cen vs info cen, bet cen vs clo cen, bet cen vs eig cen, | 11 | Nodes with high degree are mostly important nodes, having high tendency to be in the shortest path between any two nodes, having |
| | | | clo cen vs eig cen, clo cen vs info cen, eig cen vs info cen | | tendency to connect with important nodes and having more information in network, nodes having less path difference form most of the nodes are having tendency to connect with important nodes and having more information |
| bn-mouse_visual-cortex_2 | 193 | Power Law Distribution | deg cen vs pag cen, deg cen vs bet cen, deg cen vs info cen, pag cen vs bet cen, bet cen vs info cen | 5 | nodes with high degree are mostly important nodes, having tendency to be in shortest path between any two nodes and having more information |
| Component 1 network of bn-human-BNU_1_0025864_session_2-bg | 30 | Slightly Normal Distribution, Slightly Power Law Distribution | deg cen vs pag cen, deg cen vs bet cen, deg cen vs clo cen, deg cen vs eig cen, deg cen vs info cen, pag cen vs bet cen, pag cen vs clo cen, pag cen vs eig cen, pag cen vs info cen, clo cen vs eig cen, clo cen vs info cen, eig cen vs info cen | 12 | Nodes with high degree are mostly important nodes which are inclined to have less path difference form most of the nodes, having high tendency to be in the shortest path between any two nodes, having tendency to connect with important nodes and having more information in network |
| Component 2 network of bn-human-BNU_1_0025864_session_2-bg | 60 | Slightly Power Law Distribution | deg cen vs pag cen, deg cen vs eig cen, deg cen vs info cen, clo cen vs info cen | 4 | nodes with high degree are mostly important nodes, having tendency to connect with important nodes, having more information, nodes having less path differences from most other nodes |

| | | | | | are having more information |
|---|---|---|---|---|---|
| Component 3 network of bn-human-BNU_1_0025864_session_2-bg | 253 | Power Law Distribution | deg cen vs pag cen, deg cen vs eig cen, clo cen vs info cen | 3 | nodes with high degree are mostly important nodes, having tendency to connect with important nodes, nodes having less path differences from most other nodes are having more information |

Table 19: Statistical Comparison Chart

## V. Conclusion

Most of the brain network's degree distributions are following power law degree distributions while few of them have tendency to follow normal degree distribution. Most of the brain network's degree distributions have a tendency to follow more power law degree distributions as per the increasing number of nodes in the brain network. The Brain networks with large number of nodes have tendency to have less number of high correlation coefficients among these six selected network measurement algorithms. Human brain networks are following more power law degree distributions and having less number of high correlation coefficients among these six selected network measurement algorithms as per increasing number of nodes in components.

In future research, instead of using correlation methods, supervised learning [17] like a regression model [18] or artificial neural network [19] model can be used to find out the accuracy among these relationships of network measurement algorithms for brain networks. Besides brain networks, the method that is used in this paper can be applicable to networks from any discipline.